# CALCULATION OF THERMOSTABLE DIRECTIONS AND THE INFLUENCE OF BIAS ELECTRIC FIELD ON THE PROPAGATION OF THE LAMB AND SH WAVES IN LANGASITE SINGLE CRYSTAL PLATES

S.I. Burkov[1], O.P. Zolotova[1], B.P. Sorokin[2], P.P. Turchin[1], K.S. Aleksandrov[3]

*1) Siberian Federal University, 79 Svobodny ave., 660041 Krasnoyarsk, Russia*
*2) Technological Institute for Superhard and Novel Carbon Materials, 7a, Centralnaya str., Troitsk, Moscow region, Russia*
*3) L.V. Kirensky Institute of Physics, Akademgorodok, Krasnoyarsk, Russia*

e-mail: sergbsi@gmail.com; bpsorokin2@rambler.ru

***Abstract—*** **Paper is presented the results of computer simulation. Effect of the dc electric field influence on the propagation of Lamb and SH waves and its temperature coefficients of delay in piezoelectric langasite crystal plate for a lot of cuts and directions have been calculated. There were found the cuts possessing the thermostability and sufficient electromechanical coupling.**

## INTRODUCTION

Thermostability is one of the more important conditions which are expected for modern acoustoelectronic devices (resonators, narrow band filters, delay lines etc.). Langasite single crystals (LGS, $La_3Ga_5SiO_{14}$) and its isomorphs are of great importance for acoustoelectronic devices producing because of the combination of practically important properties such as thermostability, small attenuation of acoustic waves and a good electromechanical coupling in comparison with quartz [1, 2]. Application of piezoelectric crystals having the acoustic modes with low temperature coefficients and a small level of the elastic and other kinds of nonlinearities should be a good practice. So there are such devices as resonators using bulk (BAW) or surface (SAW) acoustic waves [3-8]. In addition langasite crystals possess a good chemical stability and a lack of any phase transitions up to melt point (1470 °C). Last circumstance gives a possibility of high temperature devices fabrication [9-13]. On other hand, the possibility to control and change the electromechanical properties of piezoelectric crystal by external influences such as an electric field or mechanical pressure gives us a potential to produce the controlling devices for example, the different types of acoustoelectronic sensors. As a rule, the best results for acoustoelectronic sensors can be obtained if it will be found an optimum combination of three factors: sensitivity, thermostability and good electromechanical coupling. For example, a general approach to the analysis of the acceleration sensitivity of SAW resonators has done in the paper [14] on the basis of the perturbation theory of acoustic wave's propagation under the influence of finite bias fields, given by H.F. Tiersten [15]. Some applications of this theory for purposes of experimental investigations of the complete set of nonlinear electromechanical properties of sillenite and langasite piezoelectric crystals have made by authors [16, 17]. The influence of the dc electric field, mechanical stress and temperature variation on BAW and SAW propagation in langasite single crystals has been considered in [18, 19]. Calculation of the dc electric field influence on the Lamb and shear horizontal (SH) wave propagation in $Bi_{12}GeO_{20}$ piezoelectric crystal has made earlier [20]. In present paper the anisotropy of zero-order Lamb and SH wave parameters under the action of dc electric field and the thermostable cuts and directions of these waves in langasite crystals have been investigated.

## THEORY OF LAMB AND SH WAVES PROPAGATION IN PIEZOELECTRIC PLATE UNDER THE INFLUENCE OF HOMOGENEOUS DC ELECTRIC FIELD

Influence of homogeneous dc electric field E on Lamb and SH wave propagation conditions in piezoelectric crystalline plate has been considered on the basis of the theory of bulk acoustic waves propagation in piezoelectric crystals subjected to the action of a bias electric field [16, 17]. Wave equations and electrostatics equation written in the natural state for homogeneously deformed crystals without center of symmetry have the form:

$$\rho_0 \ddot{\widetilde{U}}_i = \widetilde{\tau}_{ik,k},$$
$$\widetilde{D}_{m,m} = 0. \tag{1}$$

Here $\rho_0$ is the density of crystal taken in non-deformed (initial) state, $\widetilde{U}_i$ is the vector of dynamic elastic displacements, $\tau_{ik}$ is the tensor of thermo-dynamical stresses and $D_m$ is the vector of the induction of electricity. Here and further the tilde sign is marked the time dependent variables. Comma after the lower index denotes a spatial derivative and Latin coordinate indexes are changed from 1 to 3. Here and further the summation on twice recurring lower index is understood.

State equations can be written as:

$$\widetilde{\tau}_{ik} = C^*_{ikpq} \widetilde{\eta}_{pq} - e^*_{nik} \widetilde{E}_n,$$
$$\widetilde{D}_n = e^*_{nik} \widetilde{\eta}_{ik} + \varepsilon^*_{nm} \widetilde{E}_m, \tag{2}$$

where $\eta_{AB}$ is the deformation tensor and effective elastic, piezoelectric, dielectric constants are defined by:

$$C^*_{iklm} = C^E_{iklm} + \left( C^E_{iklmpq} d_{jpq} - e_{jiklm} \right) M_j E,$$
$$e^*_{nik} = e_{nik} + \left( e_{nikpq} d_{jpq} + H_{njpq} \right) M_j E, \tag{3}$$
$$\varepsilon^*_{nm} = \varepsilon^\eta_{nm} + \left( H_{nmik} d_{jik} + \varepsilon^\eta_{nmj} \right) M_j E.$$

In (3) E is the value of dc electric field applied to the crystal, $M_j$ is the unit vector of the E direction, $C^E_{iklmpq}$, $e_{nikpq}$, $\varepsilon^\eta_{nmj}$, $H_{nmik}$ are nonlinear elastic, piezoelectric, dielectric and electrostrictive constants (material tensors) respectively, $d_{jpq}$ and $e_{nik}$ are the piezoelectric tensors, $C^E_{iklm}$ and $\varepsilon^\eta_{nm}$ are elastic and clamped dielectric tensors. Accepting the effective material constants in the form (3) *ipso facto* we have taken into account the so-called physical nonlinearity of the piezoelectric crystalline media. Then substituting (2) into (1) we can obtain Green-Christoffel's equation in a general form which can be used for the analysis of bulk acoustic waves propagation in the case of E-influence.

Let's use coordinate system $X_3$ axis directs along the external normal to the surface of a media occupying the space $h \geq X_3 \geq 0$, and the wave propagation direction coincides with $X_1$ axis. Plane waves propagating in the piezoelectric plate are taken in the form:

$$\widetilde{U}_i = \alpha_i \exp[i(k_j x_j - \omega t)],$$
$$\varphi = \alpha_4 \exp[i(k_j x_j - \omega t)], \qquad (4)$$

where $\alpha_i$ and $\alpha_4$ are amplitudes of elastic wave and electric potential $\varphi$ concerned closely with the wave, and $k_j$ are components of wave propagation vector. Taking into account (2) and (3) the substitution (4) into (1) gives us a specific form equation. So if the electric field is applied to piezoelectric crystal, Green-Christoffel's equation can be written as

$$\left[\Gamma_{pq}(E) - \rho_0 \omega^2 \delta_{pq}\right]\widetilde{U}_q = 0, \qquad (5)$$

where Green-Christoffel's tensor has the form:

$$\Gamma_{pq} = (C^*_{ipqm} + 2d_{jkq}C^E_{ipkm}M_jE)k_ik_m,$$
$$\Gamma_{q4} = e^*_{imq}k_ik_m,$$
$$\Gamma_{4q} = \Gamma_{q4} + 2e_{ikm}d_{jkq}M_jk_ik_mE, \qquad (6)$$
$$\Gamma_{44} = -\varepsilon^*_{nm}k_nk_m.$$

The form (6) of the Green-Christoffel's tensor has been taken into account both the physical and geometrical nonlinearities. The last one is associated with the sample's dimensions and shape changing due to dc electric field influence by means of the direct piezoelectric effect. Propagation of acoustic waves in the piezoelectric plate under the E influence should satisfy to boundary conditions of zero normal components of the stress tensor on the boundaries "crystal-vacuum". Continuity of the electric field components which are tangent to the boundary surface is guaranteed by the condition of the continuity of the electrical potential and normal components of the electric displacement vector:

$$\tau_{3k} = 0, \quad x_3 = 0; \quad x_3 = h;$$
$$\varphi = \varphi^{[I]}, \quad x_3 < 0;$$
$$\varphi = \varphi^{[II]}, \quad x_3 > h; \qquad (7)$$
$$D = D^{[I]}, \quad x_3 < 0;$$
$$D = D^{[II]}, \quad x_3 > h.$$

Here the upper index «I» is concerned to the half-space $X_3 > h$ and index «II» – to the half-space $X_3 < 0$. Substituting the solutions (4) into equations (7) and neglecting of the terms which are proportional $E^2$ (and higher order ones), finally we have obtained the system of equations useful to analyze the change of the wave's structure arising as a consequence of crystal symmetry variation and new effective constants appearance:

$$\sum_{n=1}^{8} C_n \left( (C_{3ikl}^* + 2 d_{jkp} C_{3ipl}^E M_j \overline{E}) k_1^{(n)} \alpha_k^{(n)} + e_{k3j}^* k_k^{(n)} \alpha_4^{(n)} \right) \exp\left( i k_3^{(n)} h \right) = 0;$$

$$\sum_{n=1}^{8} C_n \left( (e_{3kl}^* + 2 d_{jkp} e_{3pl} M_j \overline{E}) k_1^{(n)} \alpha_k^{(n)} - (\varepsilon_{3k}^* k_k^{(n)} - i \varepsilon_0) \alpha_4^{(n)} \right) \exp\left( i k_3^{(n)} h \right) = 0;$$

$$\sum_{n=1}^{8} C_n \left( (C_{3ikl}^* + 2 d_{jkp} C_{3ipl}^E M_j \overline{E}) k_1^{(n)} \alpha_k^{(n)} + e_{k3j}^* k_k^{(n)} \alpha_4^{(n)} \right) = 0; \qquad (8)$$

$$\sum_{n=1}^{8} C_n \left( (e_{3kl}^* + 2 d_{jkp} e_{3pl} M_j \overline{E}) k_1^{(n)} \alpha_k^{(n)} - (\varepsilon_{3k}^* k_k^{(n)} + i \varepsilon_0) \alpha_4^{(n)} \right) = 0.$$

Here the index n = 1,...,4 corresponds to the number of one of the partial waves (4) and $C_n$ are the weight coefficients of the partial waves.

It can remember that the equations (8) were obtained at the assumption of homogeneity of applied dc electric field without taking into account the edge effects. But these equations are taken into consideration all the changes both the crystal density and the shape of crystalline sample arising as a consequence of finite deformation of piezoelectric media under the action of strong dc electric field (geometrical nonlinearity) [17].

## SEARCH OF THERMOSTABLE CUTS AND INVESTIGATION OF ANISOTROPY OF DC ELECTRIC FIELD INFLUENCE ON LANGASITE PIEZOELECTRIC PLATE WAVES PARAMETERS

Computer simulation of anisotropy of dc electric field influence on the Lamb and SH waves parameters of LGS piezoelectric plate has made on the basis of the theory given earlier [20-22] and above mentioned dispersive equations. Calculation of such wave parameters as phase velocities $v_i$ and $v_{im}$ for the free and metalized surface respectively, the square of electromechanical coupling coefficient (EMCC)

$$K^2 = 2 \frac{v_i - v_{im}}{v_i}, \qquad (9)$$

controlling coefficients of phase velocities by the action of dc electric field E

$$\alpha_{v_i} = \frac{1}{v_i(0)}\left(\frac{\Delta v_i}{\Delta E}\right)_{\Delta E \to 0} \tag{10}$$

and temperature coefficients of delay (TCD)

$$TCD_i = \alpha_{11} - \frac{1}{40}\frac{v_i(40°) - v_i(0°)}{v_i(20°)} \tag{11}$$

for a lot of crystal directions has been carried out by our software. Quantity $\alpha_{11}$ in the equation (11) represents an efficient coefficient of linear thermal expansion along the wave's propagation direction. Only zero-order modes were taken into account that corresponds to the thin plate condition. Data on material linear and nonlinear electromechanical properties and its temperature coefficients of LGS crystals were taken from [23, 24].

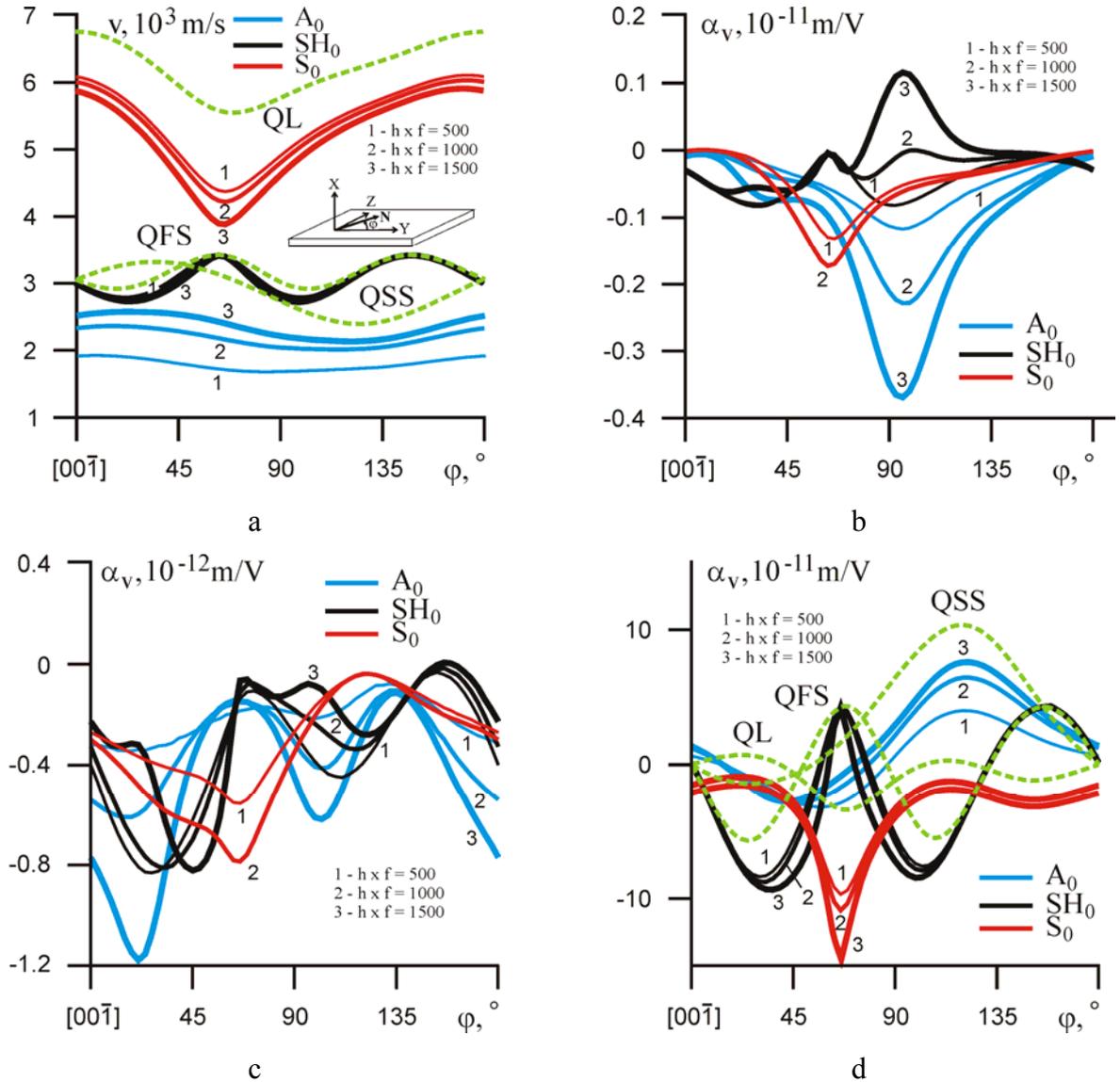

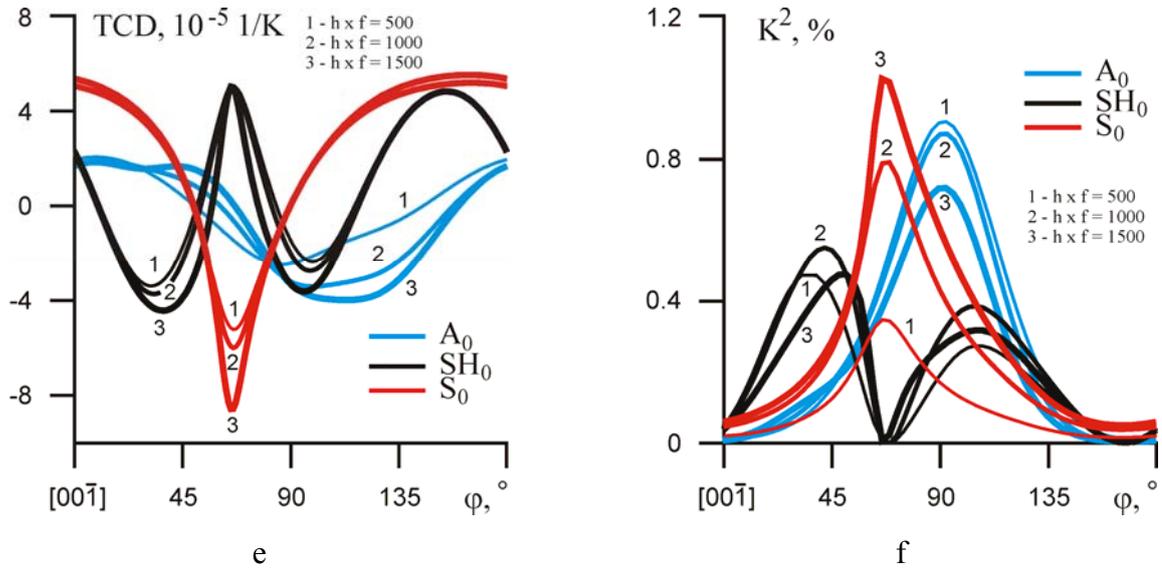

Figure 1. BAW, Lamb and SH wave's parameters for X-cut of langasite crystalline plate: a) phase velocities; b) $\alpha_v$ coefficients when E∥X₁; c) $\alpha_v$ coefficients when E∥X₂; d) $\alpha_v$ coefficients when E∥X₃; e) temperature coefficients of delay; f) electromechanical coupling coefficients. Curves for the quasi-longitudinal, fast and slow quasi-shear bulk acoustic waves are marked as QL, QFS, QSS respectively.

Wave parameters were calculated for the (100), (010), (001) crystal cuts (X, Y, and Z cuts respectively) and for a more complicated ones such as rotated Y- and Z-cuts which are of great practical importance. It was important to take into account the dispersive dependences all of the Lamb and SH wave's parameters on the h×f product where h is the plate thickness and f is the frequency. It was supposed that dc electric field was applied along $X_1$, $X_2$ and $X_3$ axes of special Cartesian coordinate system in which the $X_1$ axis coincides with wave's propagation direction $\vec{N}$ and the $X_3$ axis is parallel to the unit vector $\vec{n}$ normal to free plate surface.

<u>X cut.</u> Phase velocity values of Lamb modes are considerably changed due to h×f variation. It should be noted that the velocity of the $S_0$ symmetrical mode of Lamb wave due to h×f variation is changed oppositely to the one of the $A_0$ antisymmetrical mode. The lowest variations are observed for the $SH_0$ wave (fig. 1a). Such dependences are qualitatively the same as all the dispersive curves for Lamb waves of any materials, i. e. if h×f values are increased the phase velocities of the $A_0$ and $S_0$ modes tend to the phase velocity of Rayleigh wave from the bottom and from the top respectively.

Thermostable directions can be called the such ones in which the temperature coefficients of delay change the sign if anisotropy of TCD is calculated. Analysis of TCD anisotropy leads to the conclusion that TCD is close to the null for all the zero modes propagating along the Z axis tilted directions and lying within the sector of $\varphi = 49°…51°$ angles (fig. 1e). For example if the $S_0$ mode is taken into consideration, its parameters are: TCD = $8.1·10^{-9}$ K⁻¹ and EMCC $K^2$= 0.14 % under the orientation angle $\varphi = 49°$ and $h \times f = 500$ m/s. But there is a strong dependence between EMCC and $h \times f$ values. The $A_0$ mode has such parameters: TCD $= -9.97 \cdot 10^{-8}$ K$^{-1}$ and $K^2 = 0.21$ % under $\varphi = 50°$ and $h \times f = 500$ m/s. For the $SH_0$ wave there are TCD $= 8.65 \cdot 10^{-7}$ K$^{-1}$ and $K^2 = 0.32$ % under $\varphi = 51°$ and $h \times f = 500$ m/s. If $h \times f = 1500$ m/s (the thicker plate or higher frequency) the condition when TCD $\approx 0$ is realized for the directions $\varphi = 49° \dots 69°$: the $S_0$, $SH_0$, and $A_0$ waves have $K^2 = 0.34$ %, 0.42 %, and 0.4 % if $\varphi = 49, 57°$, and $69°$ respectively. It can be marked that there is one more direction $\vec{N} \| Y$ for the $S_0$ mode in which TCD $= 4.5 \cdot 10^{-7}$ K$^{-1}$ and $K^2 = 0.61$ % ($h \times f = 1500$ m/s).

When dc electric field coincides with wave's propagation direction the controlling coefficient of phase velocities for the $A_0$ mode has a maximal value $\alpha_v = -3.6 \cdot 10^{-12}$ m/V if $\vec{N} \| \vec{M} \| Y$ ($h \times f = 1500$ m/s). The $\alpha_v$ coefficients of the $S_0$ and $SH_0$ modes have the appreciably smaller values: $\alpha_v = -1.74 \cdot 10^{-12}$ m/V ($h \times f = 1000$ m/s) and $\alpha_v = 1.14 \cdot 10^{-12}$ m/V ($h \times f = 1500$ m/s). As a rule the absolute values of $\alpha_v$ coefficients for all the modes are increased under the $h \times f$ increasing. When dc electric field directs perpendicularly to the wave's propagation direction ($\vec{M} \| X_2 \perp \vec{N}$) the $\alpha_v$ coefficients for all the modes are considerably decreased in comparison with the $\vec{M} \| X_1 \| \vec{N}$ example. Application of dc electric field along the $X_3$ axis results in the crystalline symmetry decreasing to monoclinic one as far as the $X_3$ axis coincides with twofold symmetry axis. As a consequence the $\alpha_v$ values are considerably increased and the $S_0$ and $A_0$ modes have the maximal ones $\alpha_v = 1.5 \cdot 10^{-10}$ m/V and $\alpha_v = 7.6 \cdot 10^{-11}$ m/V respectively ($h \times f = 1500$ m/s).

As a rule the acoustoelectronics devices with the good parameters are designed using the crystalline cuts and directions with maximal electromechanical coupling. But often the directions with good EMCC have an unsatisfactory thermostability. It can be possible to compensate the temperature variations of the phase velocity and acoustic wave's path by the dc electric field applied along the direction where as a preliminary a maximal effect had found. For example the maximal value of EMCC for the $S_0$ mode makes up $K^2 = 0.89$ % (X cut, $\varphi = 63°$; TCD $= -7.74 \cdot 10^{-5}$ K$^{-1}$ and $\alpha_v = -1.365 \cdot 10^{-11}$ m/V). DC electric field influence E $= -4.03 \cdot 10^4$ V/m along $X_3$ axis gives us the TCD $= 0$ for metalized surface. The same TCD $= 0$ would be obtained by E application along $X_1$ axis but E $= -1.98 \cdot 10^6$ V/m absolute value should be considerably larger than in previous case because $\alpha_v = -2.76 \cdot 10^{-12}$ m/V.

Y-cut. The qualitative changing of the Lamb and SH waves due to $h \times f$ variation is the same as in the X-cut's case (fig. 2a). There are some possibilities where temperature coefficients of delay for Lamb and SH waves have changed a sign too. So it can see that TCD will be close to zero both for the $S_0$ mode in the angle sector $\varphi = 35°…38°$ and for the $A_0$ mode ($\varphi = 46°…68°$) if $h×f = 500…1500$ m/s (fig. 2). Particularly for the $S_0$ and $A_0$ modes there are TCD = $1.4·10^{-9}$ K$^{-1}$, $K^2 = 0.23$ % and TCD = $2.7·10^{-7}$ K$^{-1}$, $K^2 = 0.04$ % respectively if $\varphi = 39°$ and $h×f = 1500$ m/s. But the TCD of the $SH_0$ wave don't change a sign in the all observed sector of the $\varphi$ angles. When the $SH_0$ wave is propagated near $\varphi = 90^0$ direction (Z axis) there exists a minimal value of TCD = $1.51·10^{-6}$ K$^{-1}$ with $K^2 = 0.16$ % ($h×f = 500$ m/s). The $S_0$ mode has a maximal $K^2 = 2.0$ % when the wave is propagated along X axis but there is a considerable TCD = $-3.56·10^{-5}$ K$^{-1}$ value ($h×f = 500$ m/s). But the $\alpha_v$ coefficients are close to zero ($\alpha_v \sim 1·10^{-13}$ m/V) if E||$0X_3$ or E||$0X_2$, besides the E||$0X_1$ opportunity in which there is $\alpha_v = -7.13·10^{-11}$ m/V. As a consequence in this case the dc electric field E = $-8.19·10^4$ V/m should be applied along $X_1$ axis to obtain the TCD = 0 condition.

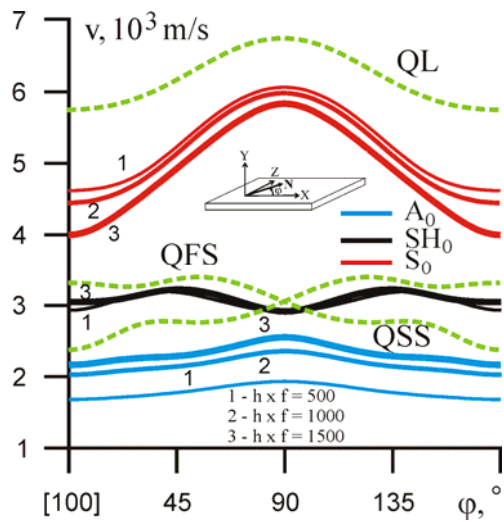

a

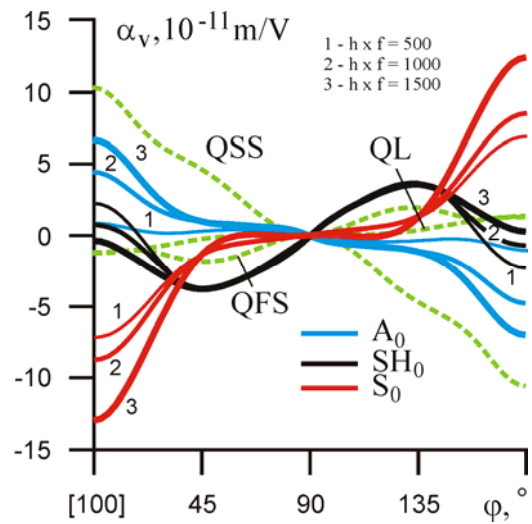

b

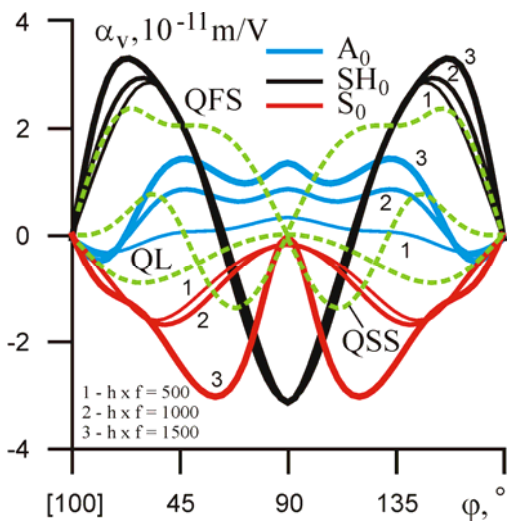

c

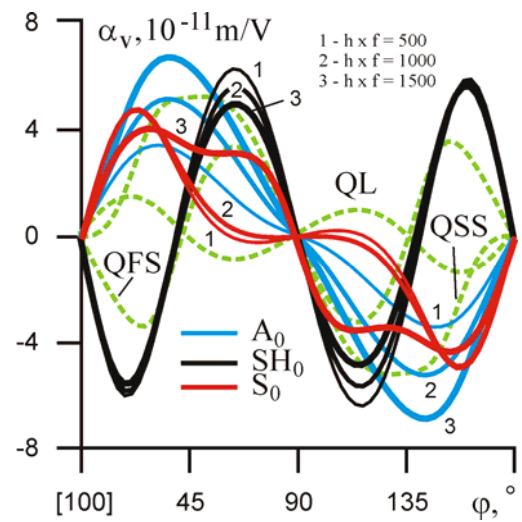

d

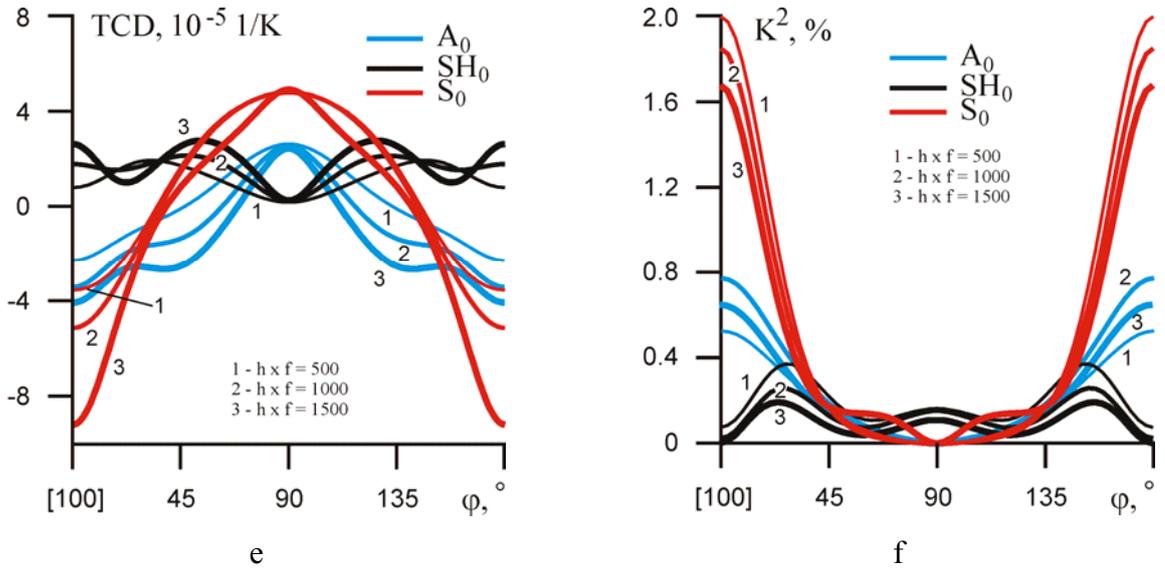

Figure 2. BAW, Lamb and SH wave's parameters for Y-cut of langasite crystalline plate: a) phase velocities; b) $\alpha_v$ coefficients when $E\|X_1$; c) $\alpha_v$ coefficients when $E\|X_2$; d) $\alpha_v$ coefficients when $E\|X_3$; e) temperature coefficients of delay; f) electromechanical coupling coefficients. Curves for the quasi-longitudinal, fast and slow quasi-shear bulk acoustic waves are marked as QL, QFS, QSS respectively.

<u>Z-cut.</u> Absence of thermostable directions for thin plates is one of the wave's propagation peculiarities of Z-cut (fig. 3). Beginning $h\times f = 1000$ m/s for the $A_0$ and $SH_0$ modes the directions possessing the change of the TCD's sign are realized. But if $h\times f$ product is increased contrariwise the EMCC values are decreased. For example there are such parameters for the $A_0$ and $SH_0$ modes propagating along the direction $\varphi = 23°$ ($h\times f = 1500$ m/s): ($A_0$) TCD = $5.9\cdot10^{-7}$ $K^{-1}$, and $K^2 = 0.17$ %; ($SH_0$) TCD = $-7.64\cdot10^{-7}$ $K^{-1}$, and $K^2 = 0.99$ %. It was taken into account that an efficient coefficient of linear thermal expansion along all the wave's propagation directions lying in Z-cut remains the constant value: $\alpha_{11} = 5.84\cdot10^{-6}$ $K^{-1}$. It should be noted such peculiarity of the wave's propagation in Z-cut: dc electric field influence along $X_1$ or $X_2$ axis has a major effect in comparison with $E\|X_3$ (note that in this case there is $X_3\|Z\|[001]$). It can be explained by the fact that the E application along Z axis coinciding with threefold symmetry axis leads to the symmetry decreasing to trigonal one. But the new effective constants induction has been produced by the nonlinear piezoelectric effect only and as a result such constants will have the small values [17].

Maximal $K^2 = 2.5$ % for the $SH_0$ mode ($h\times f = 500$ m/s) is observed in the direction $\varphi = 30°$ but TCD = $-5.41\cdot10^{-5}$ $K^{-1}$ has a considerable value. So to create a thermostable direction it is necessary to apply the dc electric field E = $-6.37\cdot10^4$ V/m along $X_2$ axis or E = $-4.67\cdot10^7$ V/m along $X_3$ axis because appropriate values are $\alpha_v = 9.16\cdot10^{-11}$ m/V and $\alpha_v = -1.26\cdot10^{-13}$ m/V respectively.

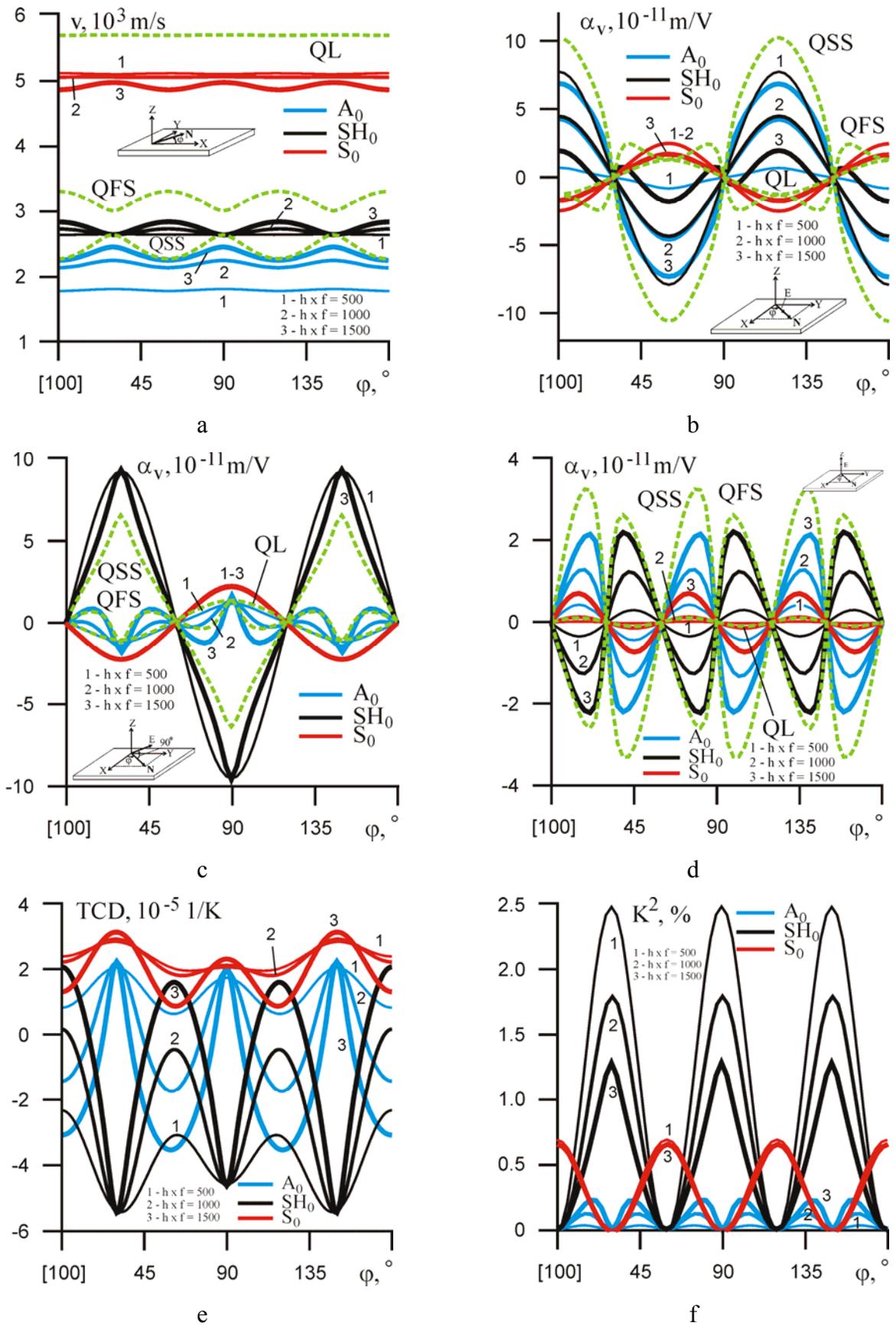

Figure 3. BAW, Lamb and SH wave's parameters for Z-cut of langasite crystalline plate: a) phase velocities; b) $\alpha_v$ coefficients when $E\|X_1$; c) $\alpha_v$ coefficients when $E\|X_2$; d) $\alpha_v$ coefficients when $E\|X_3$; e) temperature coefficients of delay; f) electromechanical coupling coefficients. Curves for the quasi-

longitudinal, fast and slow quasi-shear bulk acoustic waves are marked as QL, QFS, QSS respectively.

Rotated cuts. In practice engineers of acoustoelectronics devices have mostly used the LGS rotated cuts [25]. A distinctive peculiarity of rotated Z-cuts is the weak dependence of phase velocities and EMCC from crystalline plane's orientation. As a consequence the appropriate curves of 25°Z- and 35° Z-cuts are the same ones (fig. 4a-f). But TCD and $\alpha_v$ coefficients differ considerably (fig. 4g-r). It should be noted that the wave's parameters dependence on h×f value has a more considerable significance than the selection of crystalline plane's orientation. Thermostable directions have found for all the types of acoustic waves and for all the investigated cuts. In particular for 35° Z-cut and h×f = 1000 m/s there are such parameters of waves: $(SH_0)$ TCD = $5.8 \cdot 10^{-8}$ K$^{-1}$, and $K^2 = 0.67\%$ ($\varphi = 63°$); $(A_0)$ TCD = $1.5 \cdot 10^{-7}$ K$^{-1}$, and $K^2 = 0.24\%$ ($\varphi = 15°$); $(S_0)$ TCD = $5.44 \cdot 10^{-8}$ K$^{-1}$, and $K^2 = 0.21\%$ ($\varphi = 61°$). DC electric field influence along the wave's normal direction ($E \| X_1$) is qualitatively the same for Lamb waves (fig. 4j-l). It can be noted a minimal $E \| X_3$ influence for metalized surface on the $A_0$ phase velocity ($\alpha_v \sim 1 \cdot 10^{-12}$ m/V) (fig. 4p).

Phase velocities of the $S_0$ mode observed among a lot of the rotated Y-cuts have the highest values up to 6089.8 m/s comparable with ones which were found out among the X-cuts. A distinctive peculiarity of wave propagation for the rotated Y-cuts represents the hybridization effect, i.e. the existence of coupled acoustic modes with the energy exchange [21, 26]. For example let's take the 35° Y-cut. Such effect has arisen between $S_0$ and $SH_1$ modes (h×f = 1500 m/s) and as a result an exponential behavior of TCD и $\alpha_v$ coefficients near the (54°…84°) region has been observed (fig. 5i, l, o, r). The hybridization region is marked by vertical lines on fig. 5.

Among the rotated Y-cuts thermostable directions exist for all the modes. If we consider 15° Y-cut there is a sector of angles where TCD is close to the null for all the zero modes of the thin plate. Let's h×f = 1000 m/s. In this case the plate modes have such parameters: $(S_0)$ TCD = $6.71 \cdot 10^{-8}$ K$^{-1}$, and $K^2 = 0.29\%$ ($\varphi = 41°$); $(A_0)$ TCD = $1.9 \cdot 10^{-7}$ K$^{-1}$, and $K^2 = 0.28\%$ ($\varphi = 38°$); $(SH_0)$ TCD = $3.3 \cdot 10^{-7}$ K$^{-1}$, and $K^2 = 0.33\%$ ($\varphi = 44°$) (fig. 5). DC electric field influence has an interesting peculiarity: if $E \| X_3$, the anisotropy of $\alpha_v$ coefficients is qualitatively the same as another investigated cuts and depends more on h×f value, than on the cut's choice. But the results obtained by other orientations of dc electric field applied to 15° Y-cut sample differ considerably from previous ones (see, for example, fig. 4). If $E \| X_3$, the maximal $\alpha_v$ coefficients were found using 35° Y-cut plate (h×f = 1500 m/s): $(S_0)$ $\alpha_v = 1.4 \cdot 10^{-10}$ m/V ($\varphi = 24°$); $(SH_0)$ $\alpha_v = 8.38 \cdot 10^{-11}$ m/V ($\varphi = 57°$); $(A_0)$ $\alpha_v = -7.77 \cdot 10^{-11}$ m/V ($\varphi = 144°$). Maximal EMCC values of the $A_0$ and $S_0$ Lamb waves have a place in the $\varphi = 0°$ direction, in particular $K^2 = 0.86\%$ (35° Y-cut, h×f = 1000 m/s) and $K^2 = 1.88\%$ (65° Y-cut, h×f = 500 m/s), respectively. But the thermostability of these modes is far from desired one. Therefore if it would like to realize the TCD = 0 condition it should apply along $X_3$ axis the dc electric field to be used the values E = -1.01·10⁶ V/m and E = -2.61·10⁵ V/m, respectively. Cuts and directions of the good parameters are shown in the table. The best ones have been highlighted of the yellow color.

Table. Cuts and directions for plate waves propagation

Basic cuts

| Crystalline plane | Angle φ, deg. | Mode | Phase velocity, m/s | hf, m/s | K², % | TCD, 10⁻⁷ K⁻¹ | $\alpha_v$, 10⁻¹¹ m/V | | |
|---|---|---|---|---|---|---|---|---|---|
| | | | | | | | E‖X₁ | E‖X₂ | E‖X₃ metallized |
| | 51 | | 1757.6 | 500 | 0.21 | -9.72 | -0.04 | -0.02 | -3.07 |
| | 60 | A₀ | 2198.5 | 1000 | 0.35 | 5.66 | -0.06 | -0.02 | -2.28 |
| | 66 | | 2394.9 | 1500 | 0.37 | 8.54 | -0.11 | -0.02 | -0.87 |
| | 51 | | 3235.1 | 500 | 0.32 | 8.65 | -0.05 | -0.05 | -3.02 |
| | 117 | | 2989.1 | 500 | 0.24 | 9.67 | -0.05 | -0.04 | -4.91 |
| | 51 | SH₀ | 3208.6 | 1000 | 0.48 | -53.8 | -0.05 | -0.06 | -3.88 |
| X-cut | 117 | | 2977.8 | 1000 | 0.34 | 2.53 | -0.01 | -0.03 | -4.98 |
| | 57 | | 3300.3 | 1500 | 0.40 | -9.54 | -0.03 | -0.06 | -3.42 |
| | 117 | | 2959.2 | 1500 | 0.28 | -20.5 | 0.02 | -0.03 | -5.08 |
| | 48 | | 4827.9 | 500 | 0.15 | 27.6 | -0.06 | -0.04 | -3.23 |
| | 90 | | 4894.4 | 500 | 0.17 | 10.2 | -0.06 | -0.03 | -3.17 |
| | 48 | S₀ | 4722.6 | 1000 | 0.30 | 29.0 | -0.10 | -0.06 | -3.43 |
| | 90 | | 4807.9 | 1000 | 0.43 | 71.8 | -0.07 | -0.04 | -3.38 |
| | 51 | | 4337.8 | 1500 | 0.38 | -46.7 | -0.31 | -0.16 | -4.76 |
| | 90 | | 4605.7 | 1500 | **0.61** | 4.55 | -0.10 | -0.06 | -3.96 |
| | 48 132 | | 1791.7 | 500 | 0.13 | 7.24 | 0.32 -0.40 | 0.08 | 2.67 -2.79 |
| | 63 117 | A₀ | 2233.9 | 1000 | 0.07 | 7.1 | 0.75 -0.84 | 0.69 | 3.02 -3.17 |
| | 69 111 | | 2420.4 | 1500 | 0.04 | 2.75 | 0.61 -0.69 | 0.98 | 3.15 -3.29 |
| | 90 | | 2920.8 | 500 | 0.16 | 15.1 | -0.01 | -3.08 | 0.09 |
| Y-cut | 90 | SH₀ | 2918.4 | 1000 | 0.15 | 19.8 | -0.01 | -3.08 | -0.004 |
| | 90 | | 2917.0 | 1500 | 0.11 | 22.8 | -0.01 | -3.09 | 0.001 |
| | 36 144 | | 4980.9 | 500 | 0.36 | 14.4 | -2.27 2.20 | -1.6 | 3.21 -3.34 |
| | 36 144 | S₀ | 4889.9 | 1000 | 0.32 | -9.36 | -2.57 2.48 | -1.65 | 3.41 -3.58 |
| | 39 141 | | 4759.1 | 1500 | 0.23 | 0.014 | -2.58 2.44 | -2.0 | 3.58 -3.92 |
| | 15 165 | | 2190.2 | 1000 | 0.12 | -9.74 | 2.82 -3.12 | -0.14 | 1.26 -1.3 |
| | 45 135 | | 2190.5 | 1000 | 0.12 | -23.9 | -3.09 2.79 | -0.15 | -1.3 1.25 |
| | 75 105 | A₀ | 2190.4 | 1000 | 0.12 | -31.2 | -3.11 2.81 | -0.03 | 1.26 -1.3 |
| Z-cut | 24 156 | | 2420.8 | 1500 | 0.17 | -42.0 | 2.98 -3.34 | -0.36 | 1.97 -2.07 |
| | 36 144 | | 2421.1 | 1500 | 0.17 | -37.7 | -3.3 2.94 | -0.38 | -2.06 1.96 |
| | 84 96 | | 2420.8 | 1500 | 0.17 | 33.1 | -3.34 2.98 | -0.08 | 1.98 -2.08 |
| | 15 | SH₀ | 2744.5 | 1500 | 0.47 | -7.64 | | 4.25 | -1.98 |

| | | | | | | | | |
|---|---|---|---|---|---|---|---|---|
| | 165 | | | | | | 0.60 -0.48 | | 1.94 |
| | 45 135 | | 2745.8 | 1500 | 0.47 | 5.81 | -0.46 0.58 | 4.26 | 1.94 -1.98 |
| | 75 105 | | 2744.2 | 1500 | 0.47 | -5.01 | -0.44 0.57 | -4.28 | -1.98 1.94 |
| | 63 117 | $S_0$ | 4871.2 | 1500 | 0.65 | 87.1 | 1.66 -1.71 | 0.26 | 0.22 -0.24 |
| Y+45° -cut | 39 | $A_0$ | 1769.0 | 500 | 0.20 | -8.12 | -0.19 | -0.01 | 3.03 |
| | 39 | | 2215.9 | 1000 | 0.26 | -0.44 | -1.16 | -0.48 | 3.5 |
| | 36 | | 2405.8 | 1500 | 0.26 | -57.9 | -2.2 | -0.88 | 4.34 |
| | 45 | $SH_0$ | 3112.9 | 500 | 0.34 | 0.28 | 2.82 | -2.01 | 4.28 |
| | 159 | | 2925.5 | 1000 | 0.35 | -11.4 | -1.12 | -1.9 | 6.39 |
| | 156 | | 2948.8 | 1500 | 0.27 | 4.96 | -1.64 | -2.03 | 5.84 |
| | 174 | $S_0$ | 4868.2 | 500 | 1.0 | -7.53 | -4.7 | 1.56 | 0.59 |
| | 171 | | 4825.6 | 1000 | 0.90 | -1.15 | -5.21 | 1.84 | 0.23 |
| | 168 | | 4640.1 | 1500 | 0.78 | 0.72 | -6.58 | 2.62 | 0.61 |

## Rotated Z-cuts

| Crystalline plane | Angle $\varphi$, deg. | Mode | Phase velocity, m/s | hf, m/s | $K^2$, % | TCD, $10^{-7}$ $K^{-1}$ | $\alpha_v$, $10^{-11}$ m/V | | |
|---|---|---|---|---|---|---|---|---|---|
| | | | | | | | $E\|X_1$ | $E\|X_2$ | $E\|X_3$ metallized |
| 25° Z-cut | 36 | $A_0$ | 1757.4 | 500 | 0.28 | 7.76 | -2.61 | -0.71 | -0.09 |
| | 30 | | 2191.8 | 1000 | 0.32 | -9.47 | -3.52 | -0.39 | -1.56 |
| | 21 | | 2414.3 | 1500 | 0.26 | -3.32 | -3.83 | -0.26 | -2.30 |
| | 147 | $SH_0$ | 2898.6 | 500 | 1.4 | -14.3 | -1.64 | -5.65 | 0.10 |
| | 150 | | 2890.5 | 1000 | 0.98 | -1.37 | -1.77 | -5.39 | 0.64 |
| | 156 | | 2862.4 | 1500 | 0.70 | 2.59 | -1.92 | -5.07 | 1.29 |
| | 81 | $S_0$ | 4823.6 | 500 | 0.40 | 8.23 | -3.07 | 0.87 | 2.85 |
| | 81 | | 4729.6 | 1000 | 0.58 | -9.89 | -3.51 | 1.1 | 3.02 |
| | 84 | | 4529.2 | 1500 | 0.69 | -8.5 | -4.05 | 1.74 | 3.52 |
| 35° Z-cut | 141 | $A_0$ | 1751.1 | 500 | 0.31 | 1.18 | 2.62 | 0.64 | -0.08 |
| | 150 | | 2191.4 | 1000 | 0.32 | -2.98 | 3.43 | 0.26 | -1.57 |
| | 159 | | 2413.8 | 1500 | 0.26 | -1.67 | 3.7 | -0.11 | -2.3 |
| | 36 | $SH_0$ | 2922.3 | 500 | 1.29 | -1.74 | 2.07 | 5.27 | 0.23 |
| | 33 | | 2914.6 | 1000 | 0.91 | 5.22 | 2.1 | 5.11 | 0.68 |
| | 27 | | 2887.8 | 1500 | 0.66 | 2.04 | 2.11 | 4.92 | 1.24 |
| | 105 | $S_0$ | 4783.0 | 500 | 0.46 | -9.86 | 3.63 | -0.82 | 2.87 |
| | 102 | | 4706.5 | 1000 | 0.59 | -7.79 | 3.82 | -1.13 | 3.03 |
| | 99 | | 4503.7 | 1500 | 0.69 | -5.05 | 4.5 | -1.86 | 3.48 |
| 95° Z-cut | 33 | $A_0$ | 1762.7 | 500 | 0.25 | 0.96 | -2.5 | 0.68 | 0.04 |
| | 27 | | 2200.4 | 1000 | 0.18 | 9.77 | -3.33 | 0.3 | 1.36 |
| | 21 | | 2413.7 | 1500 | 0.11 | 0.71 | -3.87 | -0.15 | 2.15 |
| | 150 | $SH_0$ | 2871.2 | 500 | 1.51 | -23.3 | -1.2 | 5.91 | -0.05 |
| | 153 | | 2863.3 | 1000 | 1.04 | -7.52 | -1.47 | 5.65 | -0.64 |
| | 159 | | 2834.8 | 1500 | 0.73 | -0.86 | -1.79 | 5.34 | -1.37 |
| | 93 | $S_0$ | 4886.6 | 500 | 0.26 | -2.28 | -1.36 | -0.69 | -3.08 |
| | 99 | | 4809.8 | 1000 | 0.43 | -3.14 | -0.45 | -0.62 | -3.56 |
| | 96 | | 4591.6 | 1500 | 0.63 | -19.5 | -1.49 | -1.35 | -4.15 |

## Rotated Y-cuts

| Crystalline plane | Angle $\varphi$, deg. | Mode | Phase velocity, m/s | hf, m/s | $K^2$, % | TCD, $10^{-7}$ $K^{-1}$ | $\alpha_v$, $10^{-11}$ m/V | | |
|---|---|---|---|---|---|---|---|---|---|
| | | | | | | | $E\|X_1$ | $E\|X_2$ | $E\|X_3$ metallized |
| 15° Y-cut | 42 | $A_0$ | 1782.1 | 500 | 0.17 | 9.61 | 0.16 | -0.007 | 2.89 |
| | 36 | | 2203.1 | 1000 | 0.29 | -1.98 | 1.12 | 0.31 | 3.40 |
| | 36 | | 2405.3 | 1500 | 0.26 | -35.8 | 2.10 | 0.81 | 4.36 |
| | 45 | $SH_0$ | 3114.2 | 500 | 0.34 | -10.7 | -2.92 | 1.92 | 4.28 |
| | 45 | | 3105.6 | 1000 | 0.33 | -10.4 | -3.89 | 1.89 | 4.37 |
| | 42 | | 3130.9 | 1500 | 0.25 | 0.95 | -4.19 | 2.17 | 3.55 |
| | 171 | $S_0$ | 4928.7 | 500 | 0.93 | -1.03 | 4.36 | -1.56 | 0.11 |
| | 168 | | 4890.0 | 1000 | 0.81 | 3.53 | 4.78 | -1.79 | -0.13 |
| | 165 | | 4720.1 | 1500 | 0.68 | 1.84 | 5.97 | -2.38 | 0.38 |
| 35° Y-cut | 39 | $A_0$ | 1771.4 | 500 | 0.21 | 1.70 | -0.09 | -0.018 | 2.99 |

| cut | 30 | | 2196.1 | 1000 | 0.35 | -1.34 | -0.54 | -0.031 | 2.25 |
|---|---|---|---|---|---|---|---|---|---|
| | 27 | | 2413.4 | 1500 | 0.33 | 13.0 | -1.24 | -0.11 | 1.39 |
| | 42 | $SH_0$ | 3161.9 | 500 | 0.37 | 16.9 | 1.04 | -0.98 | 3.98 |
| | 42 | $SH_0$ | 3134.0 | 1000 | 0.49 | -30.8 | 1.29 | -1.01 | 4.67 |
| | 36 | | 3218.0 | 1500 | 0.42 | -14.5 | 1.75 | -1.36 | 4.03 |
| | 3 | $S_0$ | 4792.5 | 500 | 0.32 | 28.3 | 1.75 | -0.79 | 3.66 |
| | 3 | $S_0$ | 4697.9 | 1000 | 0.56 | 20.6 | 2.03 | -1.02 | 3.88 |
| | 3 | | 4471.1 | 1500 | 0.72 | 10.3 | 2.80 | -1.78 | 4.43 |
| 55° Y-cut | 42 | $A_0$ | 1774.7 | 500 | 0.17 | -8.17 | -0.30 | -0.07 | 2.99 |
| | 60 | $A_0$ | 2241.8 | 1000 | 0.09 | 8.81 | -0.91 | -0.87 | 3.18 |
| | 66 | | 2423.9 | 1500 | 0.06 | -22.2 | -0.81 | -1.29 | 3.57 |
| | 66 | $SH_0$ | 2968.5 | 500 | 0.17 | -0.08 | 2.28 | 0.78 | 6.64 |
| | 84 | $SH_0$ | 2907.1 | 1000 | 0.16 | 3.52 | 0.32 | 2.77 | 2.07 |
| | 162 | | 2993.9 | 1500 | 0.22 | 16.3 | -2.08 | -3.29 | 6.08 |
| | 153 | $S_0$ | 4939.3 | 500 | 0.69 | -10.4 | -3.78 | 1.58 | -3.39 |
| | 150 | $S_0$ | 4904.7 | 1000 | 0.45 | 13.6 | -3.80 | 1.62 | -3.18 |
| | 150 | | 4698.5 | 1500 | 0.37 | 6.89 | -5.07 | 1.61 | -2.81 |
| 65° Y-cut | 135 | $A_0$ | 1786.3 | 500 | 0.15 | -4.67 | 0.26 | -0.097 | -2.89 |
| | 120 | $A_0$ | 2242.2 | 1000 | 0.09 | -7.02 | 0.82 | -0.87 | -3.29 |
| | 111 | | 2446.2 | 1500 | 0.05 | 9.41 | 0.67 | -1.24 | -3.16 |
| | 93 | $SH_0$ | 2915.6 | 500 | 0.16 | 5.37 | 0.08 | 2.82 | -1.04 |
| | 96 | $SH_0$ | 2907.3 | 1000 | 0.16 | -5.08 | -0.35 | 2.77 | -2.08 |
| | 21 | | 3015.5 | 1500 | 0.22 | -3.86 | 2.07 | -3.21 | -5.95 |
| | 27 | $S_0$ | 4938.9 | 500 | 0.70 | -3.59 | 3.72 | 1.58 | 3.25 |
| | 30 | $S_0$ | 4904.4 | 1000 | 0.45 | 20.7 | 3.73 | 1.62 | 3.03 |
| | 30 | | 4698.4 | 1500 | 0.37 | 8.11 | 4.97 | 1.61 | 2.54 |

## CONCLUSION

Phase velocities of the $S_0$ modes in langasite have the highest values up to 6100 m/s in comparison with ones of the bulk shear waves, the $SH_0$ and $A_0$ Lamb modes. This circumstance can be taken into account in practical use. It should emphasize that on the contrary with the $A_0$ modes phase velocities of the $S_0$ modes are decreased if h×f product is increased.

Using the data of the table and fig. 1-5 it can conclude that in langasite there are some Lamb and SH modes possessing the thermostability and sufficient electromechanical coupling which would be of practical importance. Estimation of dc electric field influence indicates that E variation may cause a delay's change which can compensate the temperature fluctuations of these modes' velocities. It should be noted that thermostability of Lamb and SH modes follows a strong dependence on h×f product.

This paper was supported by grant N 4645.2010.2 (Science Schools) of President of Russia.

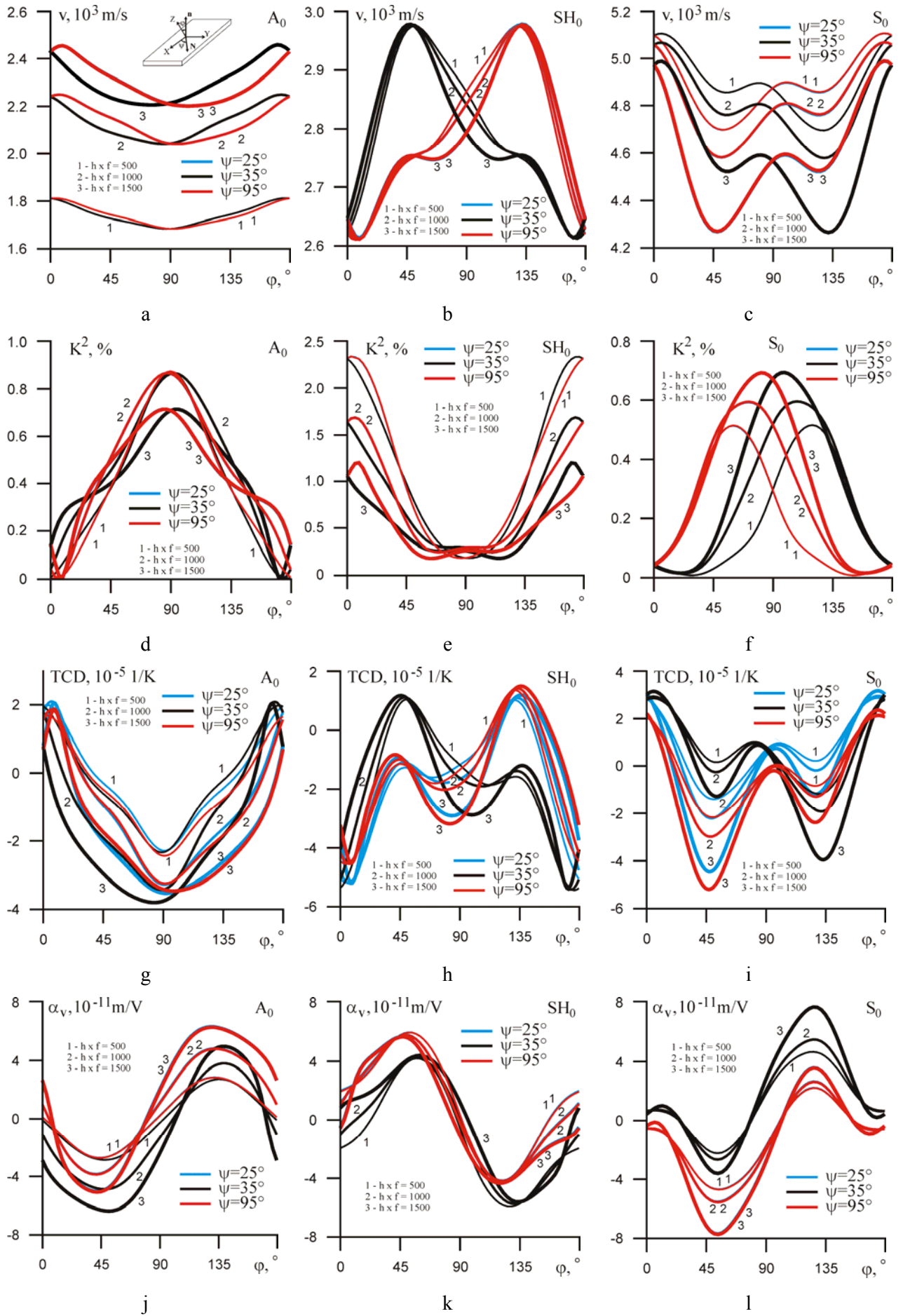

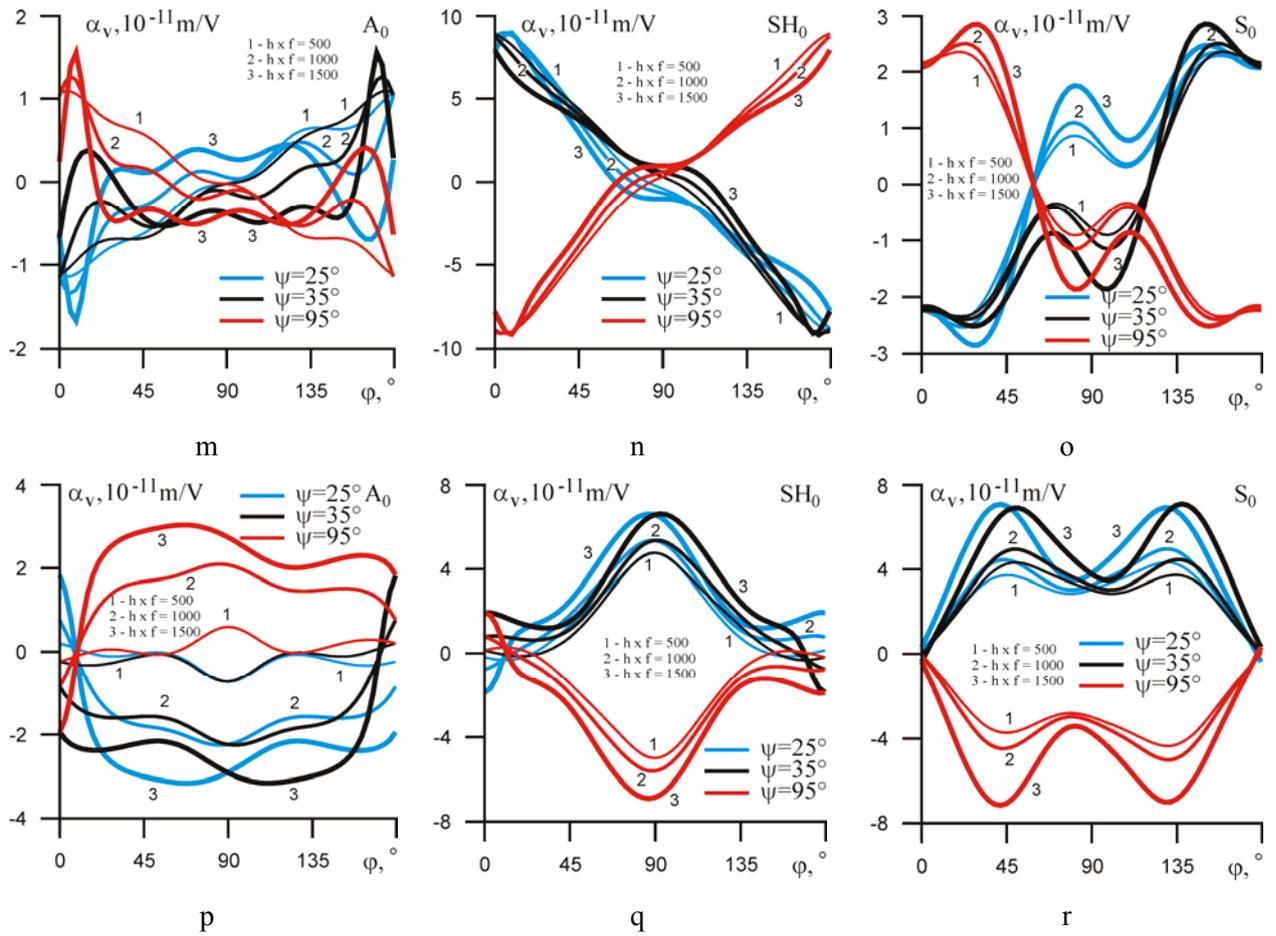

Figure 4. Zero order Lamb and SH wave's parameters for rotated Z-cuts of langasite crystalline plate: a-c) phase velocities; d-f) electromechanical coupling coefficients; g-i) temperature coefficients of delay; j-l) $\alpha_v$ coefficients when $E \| X_1$; m-o) $\alpha_v$ coefficients when $E \| X_2$; p-r) $\alpha_v$ coefficients when $E \| X_3$. A, d, g, j, m, p - $A_0$ mode; b, e, h, k, n, q - $SH_0$ mode; c, f, i, l, o, r - $S_0$ mode.

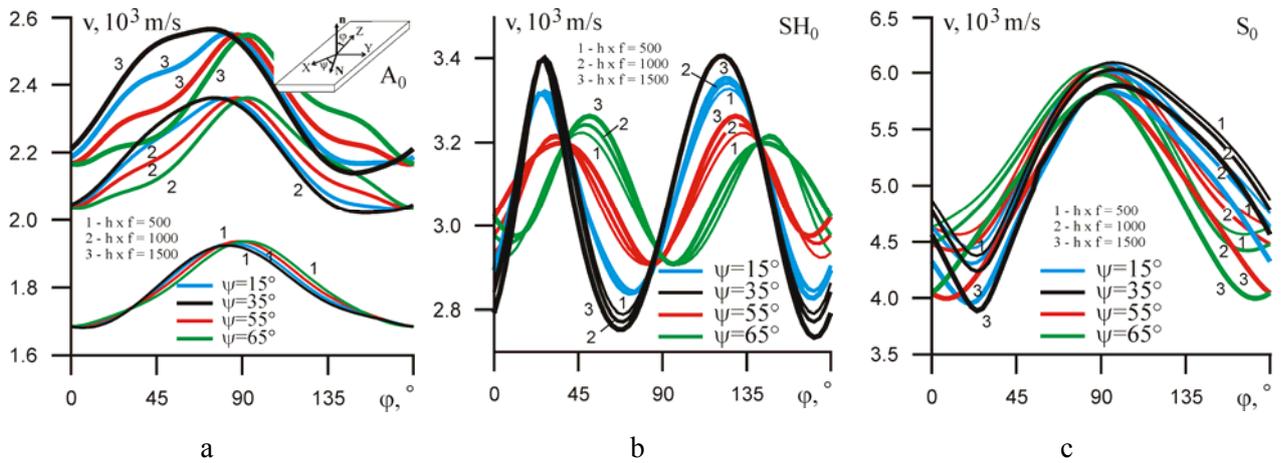

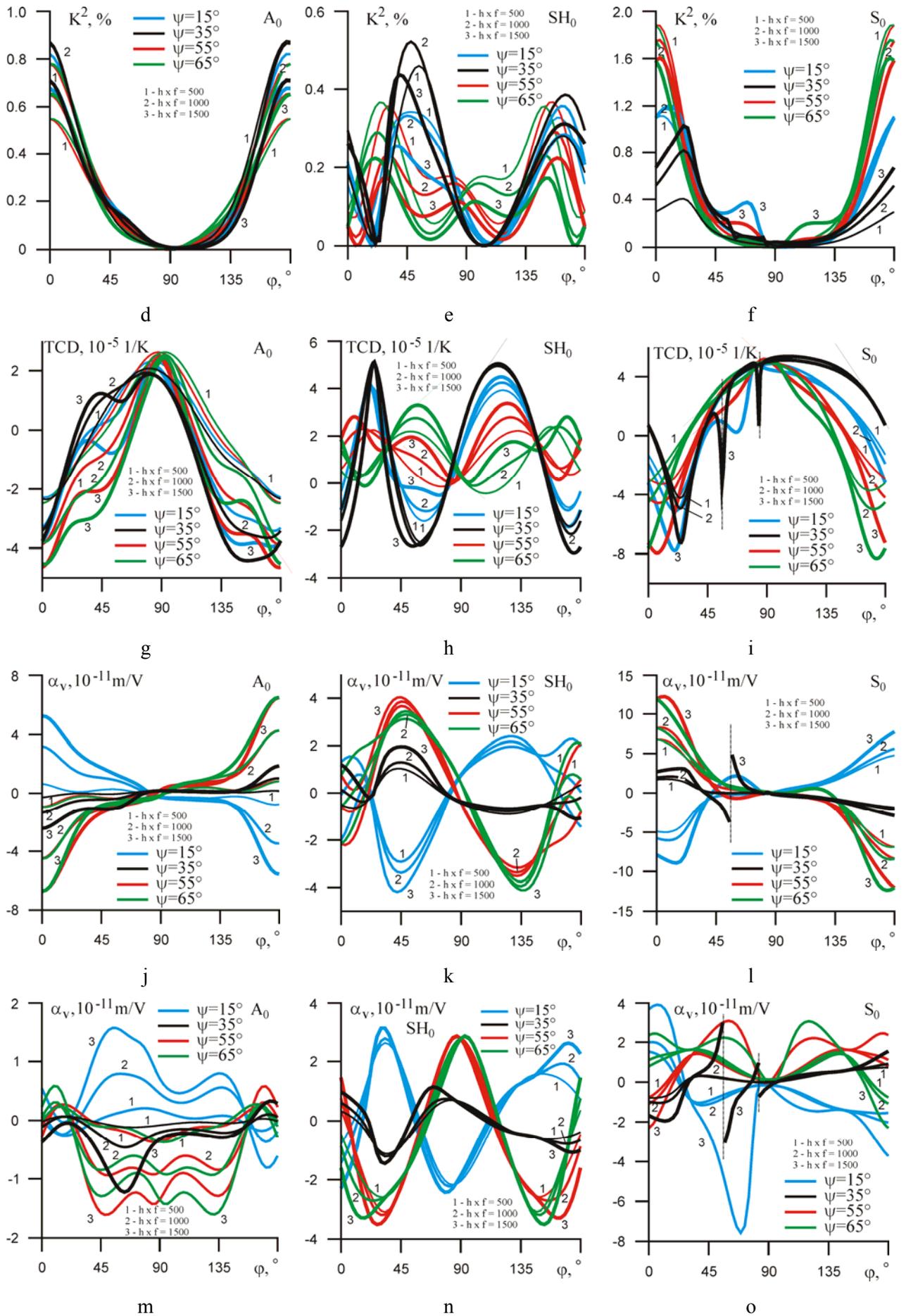

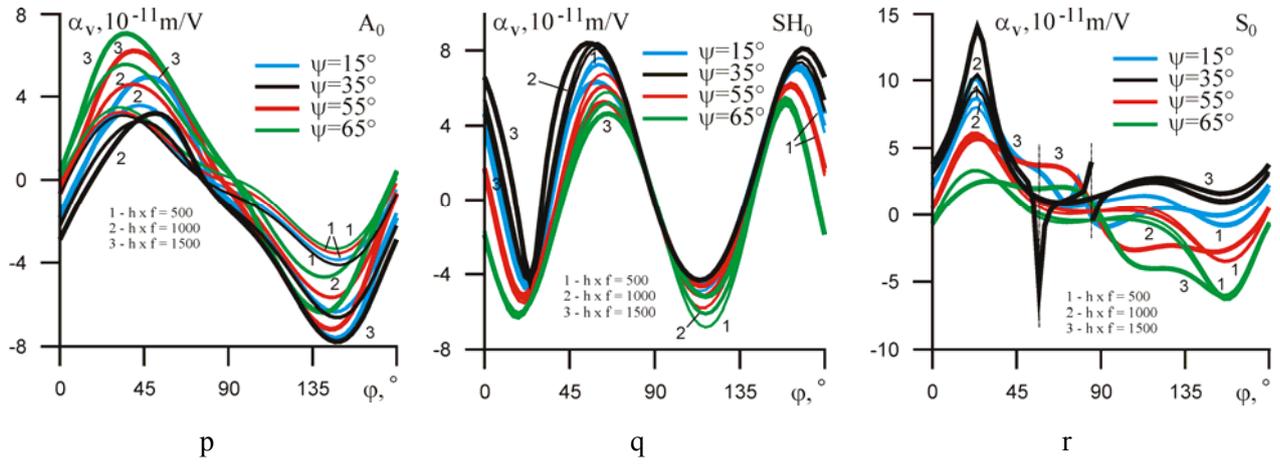

Figure 5. Zero order Lamb and SH wave's parameters for rotated Y-cuts of langasite crystalline plate: a-c) phase velocities; d-f) electromechanical coupling coefficients; g-i) temperature coefficients of delay; j-l) $\alpha_v$ coefficients when $E\|X_1$; m-o) $\alpha_v$ coefficients when $E\|X_2$; p-r) $\alpha_v$ coefficients when $E\|X_3$. A, d, g, j, m, p - $A_0$ mode; b, e, h, k, n, q - $SH_0$ mode; c, f, i, l, o, r - $S_0$ mode.